\documentclass[useAMS,usenatbib,a4]{mn2e}

\usepackage{epsfig}
\usepackage{rotating}
\usepackage{times}
\usepackage{graphicx}
\usepackage{lscape}
\usepackage{times}
\newcommand\apj{ApJ}
\newcommand\apjl{ApJL}
\newcommand\apjs{ApJS}
\newcommand\aap{A\&A}
\newcommand\mnras{MNRAS}
\newcommand\gcn{GCN Circ.}

\begin{document}
\title[The GRB Variability/Peak Luminosity Correlation]{Testing the gamma-ray burst 
variability/peak luminosity correlation on a Swift homogeneous sample}
\author[D. Rizzuto et al.]{D. Rizzuto$^{1}$,
C. Guidorzi$^{1,2}$\thanks{E-mail: cristiano.guidorzi@brera.inaf.it}, 
P. Romano$^{1,2}$, S. Covino$^{2}$, S. Campana$^{2}$,
M. Capalbi$^{3}$,\newauthor G. Chincarini$^{1,2}$, G. Cusumano$^{4}$, D. Fugazza$^{2}$,
V. Mangano$^{4}$, A. Moretti$^{2}$, M. Perri$^{3}$,\newauthor  G. Tagliaferri$^{2}$\\
$^{1}$ Universit\`a degli Studi di Milano, Bicocca, Piazza delle Scienze 3, I-20126, Milano, Italy \\
$^{2}$INAF--Osservatorio Astronomico di Brera, Via E.\ Bianchi 46, I-23807 Merate (LC), Italy \\
$^{3}$ASI Science Data Center, via G. Galilei, I-00044 Frascati (Roma), Italy\\
$^{4}$INAF--Istituto di Astrofisica Spaziale e Fisica Cosmica Sezione di Palermo, via U. La Malfa 153, I-90146 Palermo, Italy}

\date{}


\maketitle

\label{firstpage}
\begin{abstract}
We test the gamma-ray burst correlation between temporal variability and peak
luminosity of the $\gamma$-ray profile on a homogeneous sample of 36 {\it Swift}/BAT
GRBs with firm redshift determination.
This is the first time that this correlation can be tested
on a homogeneous data sample. The correlation is confirmed, as long as the 6 GRBs
with low luminosity ($<5\times10^{50}$~erg~s$^{-1}$ in the rest-frame 100-1000~keV
energy band) are ignored. We confirm that the considerable scatter of the
correlation already known is not due to the combination
of data from different instruments with different energy bands, but it is
intrinsic to the correlation itself.
Thanks to the unprecedented sensitivity of {\it Swift}/BAT,
the variability/peak luminosity correlation is tested on low-luminosity GRBs.
Our results show that these GRBs are definite outliers.
\end{abstract}



\section[]{Introduction}
A number of correlations between intrinsic properties of Gamma-Ray Bursts
(GRBs) has been discovered since it has become possible to measure their
distances. In particular, correlations between properties of the $\gamma$-ray
prompt emission as well as of the afterglow at different wavelengths
have provided an increasing number of clues to identify the mechanisms
and, ultimately, the nature of the GRB progenitors.
In addition, some of these correlations have been tentatively used
as luminosity estimators, with several implications on their possible
usage to constrain the cosmology of the Universe \citep{Ghirlanda04_cosmology,
Liang05_relation,Firmani05}.

The increasing number of GRBs with spectroscopic redshift allows
to test and better calibrate them. Recently, a crucial contribution
has been supplied by the {\em Swift} satellite \citep{Gehrels04:SWIFT},
whose average rate of 100~GRBs per year since launch (November 2004)
made it possible to measure the distances of almost 1/3 of its
sample, thus duplicating the overall number of GRBs with known redshift
since 1997. 

The sample of GRBs detected with the {\it Swift}
Burst Alert Telescope (BAT; Barthelmy et al. 2005)\nocite{Barthelmy2005:BAT}
is particularly suitable to test the correlations between intrinsic
properties, with the unprecedented benefit of a homogeneous data set,
apart from those requiring the peak energy measurement, made difficult
by the limited energy band (15--350~keV).

Hereafter we focus on a long-standing correlation
between the variability and peak luminosity of the $\gamma$-ray prompt
emission \citep{Fenimore00,Reichart2001:luminosity}.
In particular, \nocite{Reichart2001:luminosity} Reichart et~al. (2001;
hereafter R01) provided
a definition of variability (hereafter denoted as $V_{\rm R}$)
that turned out to correlate with the
isotropic-equivalent rest-frame 100--1000~keV peak luminosity (hereafter
$L$) for a sample of 11 GRBs with known redshift available at the time,
using data from the {\it CGRO}/BATSE experiment \citep{Paciesas99}.

R01 modelled the variability/peak luminosity correlation (hereafter 
$V/L$ correlation) with a power law ($L\propto\,V_{\rm R}^{m}$) with $m=3.3^{+1.1}_{-0.9}$)
affected by {\em extrinsic} or {\em sample} scatter, described by $\sigma_{\log{V_{\rm R}}}=0.18$.
Recently,\nocite{Guidorzi2005:var_peaklum} Guidorzi et~al. (2005; hereafter GFM05) and
\nocite{Guidorzi2005b} Guidorzi (2005; hereafter G05) tested the $V/L$ correlation
on an extended sample of 32 GRBs with known redshift (GFM05) and on
551 BATSE GRBs, respectively.
For the latters, a pseudo-redshift was derived assuming the lag-luminosity
correlation \citep{Norris00,Band04}.

Both works confirmed the correlation, but with a lower slope than that derived by R01:
$m = 1.3_{-0.4}^{+0.8}$ (GFM05) and $m = 0.85\pm 0.02$ (G05).
However, in either case it was pointed out that the scatter around these
power laws made the description of a simple power law unsatisfactory.
\citet{Reichart05} applied the same method as R01 to the very results obtained
by GFM05, obtaining $m = 3.4_{-0.6}^{+0.9}$ and $\sigma_{\log{V_{\rm R}}}=0.20\pm0.04$,
perfectly in agreement with the original values of R01.
They ascribed the disagreement to the fact that GFM05 did not deal with the
sample variance properly.

More recently, \citet{Guidorzi2006a} applied the D'Agostini (2005) method,
accounting for the sample variance, to the data sets of both GFM05 and G05.
They obtained shallower slopes than those by R01 and \citet{Reichart05}
and larger scatters: in particular, for the sample of 32 GRBs with firm
redshift drawn from GFM05 they obtained $m=1.7\pm0.4$, $\sigma_{\log{V_{\rm R}}}\sim0.34$,
while for the sample of 551 GRBs with pseudo-redshifts of G05 it
resulted $m=0.88_{-0.13}^{+0.12}$, $\sigma_{\log{V_{\rm R}}}\sim0.74$.

For more details on the debate concerning the methods to be used, we refer
the reader to the original papers by \citet{Reichart05} and \citet{Guidorzi2006a}.

Li \& Paczy{\'n}ski\nocite{Li2006:var_lum} (2006; hereafter LP06) have recently provided a slightly
modified definition of variability, hereafter denoted as $V_{\rm LP}$, which
they found to correlate more tightly with $L$ than $V_{\rm R}$, without any
extrinsic scatter in addition to the uncertainties affecting the single values
of the single GRBs. $V_{\rm LP}$ differs from $V_{\rm R}$ mainly in the
choice of the smoothing filter determining the reference light curve with
respect to which the variance is evaluated. LP06 chose the Savitzy-Golay
filter instead of a simple boxcar used by R01. As a result, $V_{\rm LP}$
selects only the high frequencies, whereas only in the $V_{\rm R}$
calculation the lower frequency variance can give a contribution.

The variability of the $\gamma$-ray prompt emission light curves is
supposed to be produced above the photospheric radius of the fireball,
above which radiation becomes optically thin. The interpretations proposed
of the $V/L$ correlation mainly invoke the presence of a jet, whose
angle $\theta$, i.e. either the opening angle or the viewing angle
(e.g., see Ioka \& Nakamura 2001)\nocite{Ioka01}
for some jet patterns, is strongly connected with the observed peak
luminosity $L$ as well as with the Lorentz factor $\Gamma$ of the
expanding shell(s). The result would be a strong dependence of
both $L(\theta)$ and $\Gamma(\theta)$ on $\theta$.
For instance, \citet{Kobayashi02} reproduced the observed correlation
through numerical simulations, assuming $\Gamma\propto\theta^{-q}$
and a log-uniform distribution in the time delay between next shells,
from 1~ms to 1~s. A value of $q=2$ seems to account well for the
results by \citet{Guidorzi2006a} as well as the anti-correlation between
break time and peak luminosity \citep{Salmonson02}.
Similar results have been found by \citet{Meszaros02} and
\citet{Ramirez02} under slightly different assumptions.

The new piece of information from this analysis is given
by the presence of low-luminosity high-variability GRBs.

In this paper, we test the $V/L$ correlation on a homogeneous sample
of 41~GRBs detected with {\em Swift}/BAT using fully homogeneous data.
We considered two different definitions of variability: that by R01
and that by LP06.
In Section~\ref{s:sample} we describe the data sample and the selections
we made. Sections~\ref{s:lum_peak} and \ref{s:V} report how peak luminosity
and variability have been calculated. Results are reported in
Sec.~\ref{s:res} and discussed in Sec.~\ref{s:disc}.

\section[]{The GRB sample}
\label{s:sample}

The sample includes 51 long ($T_{90} >2$s) GRBs with spectroscopic redshift detected by {\it Swift}/BAT
\citep{Gehrels04:SWIFT} between the launch (2004, November 20) and October 2006. 
Out of this sample we selected only those bursts whose $\gamma$-ray profile is entirely
covered by BAT during the burst mode \citep{Barthelmy2005:BAT}.
No further selection was made on the sample, in order to avoid any arbitrary bias in the results.
This requirement resulted in the rejection of 10 GRBs. In fact,
in these cases the observation of  {\it Swift}/BAT
switched from burst mode to the survey mode before of the end of the prompt emission.
The light curve results with a truncated profile. This is the case of GRB~050318,
whose light curve stops about 32~s after the trigger, as well as of GRB~050820A, GRB~050904 and 
GRB~060218. For GRB~060124 only the precursor was recorded in event mode, while the main event
was observed in survey mode. 
For GRB~060906 the light curve is incomplete at the beginning,
because the trigger probably missed the true onset of the burst.
No burst mode event file is available for GRB~060505, as BAT observed it only
in survey mode.
We chose not to make use of the background subtracted light curves acquired during the survey
mode to keep the sample as homogeneous as possible.
GRB~050408 was detected by XRT and UVOT, but not by BAT, although
the light curve of its prompt emission is available from other instruments
({\it HETE-2/FREGATE}; Atteia et~al. 2003).\nocite{Atteia03}
Nevertheless, we did not consider it in this work because we focused on BAT data for the
reasons reported above.
In the case of GRB~050802 and GRB~051227A the problem is in the redshift determination.
For the former only a tentative redshift exists \citep{Cummings05_GCN3479}, which
is at odds with the interpretation of the {\it Swift}/UVOT results \citep{McGowan05_GCN3745}.
For GRB~051227A there is a redshift determination of the putative host galaxy \citep{Foley05_GCN4409},
but it is still unclear if this is the real host galaxy.

After this selection the sample has shrunk to 41 long GRBs, entirely covered by BAT and processed
through the same procedure. Therefore, this work investigates the $V/L$
relation based on a completely homogeneous sample.

The BAT event files were retrieved from the {\it Swift} public archive
\footnote{http://swift.gsfc.nasa.gov/docs/swift/archive/}
and analysed through the standard BAT analysis software distributed within FTOOLS v6.1.
For each GRB we extracted mask-tagged light curves  for a number of different
binning times in the total nominal energy band ($15$--$350$~keV)
\footnote{The effective band is $15$--$150$~keV, because photons with energy above $150$~keV become
transparent to the coded mask and are treated as background by the mask-weighting technique
(e.g., Sakamoto et al. 2006).\nocite{Sakamoto06}}, through the tool {\em batmaskwtevt}
adopting the ground-refined coordinates provided by the BAT team for each burst.
These curves are therefore already background subtracted according to the
coded mask technique (Barthelmy et al. 2005 \nocite{Barthelmy2005:BAT} and references therein).
For each burst the BAT detector quality map was obtained by processing the next earlier enable/disable
map of the detectors, telling which detectors were disabled in flight because too noisy.
We also applied the energy calibration to the event file making use of the closest-in-time gain/offset
file through the tool {\em bateconvert}, as suggested by the BAT 
team\footnote{http://swift.gsfc.nasa.gov/docs/swift/analysis/threads.}.
Finally these light curves are expressed as count rates with uncertainties: the rates are
background-subtracted counts per second per fully illuminated detector for an equivalent on-axis source,
as the default corrections are applied: {\em ndets, pcode, maskwt, flatfield}.

We also studied the behaviour of the background fluctuations in burstless regions of the light curves
and we found that the mask-tagged rates, $r_i$, fluctuate compatibly with a white noise with
sigma $\sigma_{r_i}$ ($r_i$ and $\sigma_{r_i}$ are
the rate and its uncertainty of the $i$-th bin, respectively; see Appendix).
We concluded that an upper limit of $\sim$~2--4\% (4--6\%) at 90\% (99\%) confidence level can be
derived on the presence of a possible extra variance (of instrumental origin, for instance) in addition
to that due to the Poisson counting statistics, implicitly assumed during the light curve extraction
with the tool {\em batbinevt}.

We found that it is not correct to perform the same analysis on BAT light curves with raw counts,
i.e. not masked. In fact,
we found that the GRB profile itself can be dramatically contaminated by other sources and by background
variations, with time, due to the slewing of the spacecraft during the prompt emission, for most GRBs.
Furthermore, we found that BAT light curves with raw counts are severely affected by extra variance,
which is comparable with the Poisson variance due to the counting statistics, in agreement with previous
results (LP06). Therefore, we conclude that the BAT light curves of most GRBs with raw counts, not masked,
are not suitable for temporal variability studies.

\section[]{Peak luminosity}
\label{s:lum_peak}
For each GRB we extracted the mask-tagged light curve with a binning time of $50$~ms in the 15--350~keV energy
band. We determined the 1-s time interval with the highest total counts and assumed this as the time interval
corresponding to the 1-s peak count rate.

We extracted the mask weighted spectrum in this time interval using the tool {\em batbinevt}.
We applied all the corrections required: we updated it through {\em batupdatephakw} and generated
the detector response matrix using {\em batdrmgen}. Then we used {\em batphasyserr} in order to account
for the BAT systematics as a function of energy.
Finally we grouped the energy channels of the spectrum by imposing a 5-$\sigma$ (or 3-$\sigma$ when the S/N
was too low) threshold on each grouped channel.
We fitted the resulting photon spectrum, $\Phi(E)$ (ph cm$^{-2}$s$^{-1}$keV$^{-1}$),
with a power law with pegged normalisation ({\sc pegpwrlw} model under XSPEC v.12), except for
GRB~050525A and GRB~060927 where a cutoff power law was used, in the rest-frame
energy band $100-1000$~keV. 
The choice of the energy band is connected with the original
definition by R01 (see their eq.~9) also used by GFM05 (their eq.~7).

Therefore the GRBs rest-frame $100-1000$-keV isotropic-equivalent peak luminosities were computed
using:
\begin{equation}
\displaystyle L \ =\  4 \pi D_L^2(z)\ \int_{100/(1+z)}^{1000/(1+z)} E \, \Phi(E)\,dE
\label{eq:lum}
\end{equation}
where $D_L(z)$ is the luminosity distance at redshift $z$, $E$ is energy expressed in keV.
Finally we derived the uncertainty on the peak luminosity by propagating that of the measured
flux. 

Concerning the six BAT GRBs shared with the sample of GFM05, we compared the two sets of
peak luminosities: these GRBs are 050315, 050319, 050401, 050505, 050525A and 050603. 
They are consistent with those of GFM05, apart from two cases.
For 050401 our $L_{50}$ measure,
where $L_{50}=L/(10^{50}~{\rm erg~s}^{-1})$, is $1405\pm165$, while GFM05 reported
$740\pm100$. For 050603, we obtained $L_{50}=2706\pm1470$ to be compared with GFM05's
$1200\pm300$. The reason in either case resides in a slightly different choice of the
1-s time interval around the peak. GFM05 determined this from the 40--350~keV light curve
to match the 40--700~keV of the {\it BeppoSAX}/GRBM, while we used the 15--350~keV.
The choices of the 1-s time interval turned out to differ by 1--2~s in either case.
This, combined with the fact that both of these GRBs exhibit a sharp peak, turned into
the discrepancies provided above.
We note that in both cases they still lie in the $V_{\rm R}$-$L$ region
with high $V_{\rm R}$ and high $L$, consistently with the $V/L$ correlation.

\section[]{Variability}
\label{s:V}

\subsection[]{R01 definition}
\label{s:V_R01}
The main difference between our data set and those used by R01 and GFM05 is that
our light curves are expressed in background-subtracted rates and not in counts. 
This fact is due to the way BAT, which is a coded mask, has been conceived.
Hereafter we assumed a Poissonian variance for the statistical fluctuations of the light
curves, as we proved in Appendix A.
The formula we used to compute the variability, according to the R01 definition,
is basically the same as those of R01 and GFM05, with no extra-Poissonian noise term,
given that our rates are already background-subtracted.
\begin{equation}
  V_{\rm R} = \frac{\sum_{i=1}^N
    [(\sum_{j=1}^Na_{ij}r_j)^2-\sum_{j=1}^Na_{ij}^2\,\sigma_{r_j}^2]}{\sum_{i=1}^N
    [(\sum_{j=1}^Nb_{ij}r_j)^2-\sum_{j=1}^Nb_{ij}^2\,\sigma_{r_j}^2]}
  \label{eq:var_R01}
\end{equation}
where $a_{ij}$ and $b_{ij}$ are the same coefficients as those introduced by R01 in their eqs.~6-7.
The differences between our formula, eq.~7 of R01 and eq.~4 of GFM05, are
the replacement of the counts $C_j$ with the rates $r_j$ in the first terms of both
numerator and denominator, where the original $C_j$ represented the GRB signal,
and the replacement of the counts $C_j$ with the statistical noise variances $\sigma_{r_j}^2$
affecting the rates $r_j$ in the terms to be subtracted, where the original counts $C_j$
represented the noise. The sum, $j=1,\ldots,N$, runs over the $N$ bins encompassing the GRB time profile.
The background term $B_j$ in the original formulae of R01 has been set to zero, as it has already
been removed during the extraction of the light curves.

For each GRB we estimated the smoothing time scale $T_f$ ($f=0.45$), defined by R01 as the shortest cumulative
time interval during which a fraction $f$ of the total counts above background has been collected. 
For each GRB we calculated $T_f$ and the corresponding variability $V_{\rm R}$ as a function of the binning time.
We chose the values obtained with the binning time $\Delta\,t$ that fulfilled the requirements reported by GFM05
concerning the ratio $\Delta\,t/T_f$. On one side, when this ratio is too small, the light curve is dominated by
statistical fluctuations, while, on the other side, when the binning is too coarse the variability is underestimated.
A detailed description of these criteria is provided by GFM05.

\subsection[]{LP06 definition}
\label{s:V_LP06}
Concerning the definition of variability given by LP06, hereafter denoted by $V_{\rm LP}$,
we point out a number of different choices with respect to the their analysis.
First we estimated $V_{\rm LP}$ from the background-subtracted mask-tagged light curves,
while LP06 used the raw counts light curves of the 7 {\it Swift}/BAT GRBs of their sample
(Li, private comm.). We assumed no extra-Poissonian variance to be subtracted, unlike
LP06. We adapted eqs.~1--3 of LP06 accordingly and obtained the following:
\begin{equation}
  V_{\rm LP} = \frac{\sum_{i=1}^{N} \left[ W\,(r_i -y_i)^2
                - \sigma^2_{r_i}\right] }{(N-1)\, r_{\rm max}^2}
  \label{eq:var_LP06}
\end{equation}
where $y_i$ is the value for the $i$-th bin of the reference light curve obtained with
the Savitzky-Golay filter with a smoothing window of T$_f$ ($f=~0.45$).
$W$ is the same weight as that used by LP06 and accounts for the fact that the set
of ${y_i}$ is not completely statistically independent from ${r_i}$. 
As for the determination of the peak count rate, $r_{\rm max}$, we searched the
light curve of the same GRB a number of times, each time increasing the binning time,
until we found the peak 5-$\sigma$ higher than the contiguous bins. This turned out
to be very accurate, particularly for weak GRBs.
In order to comply with the procedure of LP06, $N$ corresponds to the total number of
bins encompassing the time interval which defines the $T_{90}$, i.e. from 5\% to 95\% of the
total fluence. The values of $T_{90}$ have been calculated using the ftool {\em battblocks}.
Values of $V_{\rm LP}$ have been derived from the 64-ms light curves.

\section[]{Results}
\label{s:res}
Table~\ref{table:results} reports the results of $V_{\rm R}$, $V_{\rm LP}$, 
$L$ and $T_{f=0.45}$ obtained for the sample of 41~GRBs.

 \begin{table*}
 \begin{center}
 \caption{Variability, according to both definitions considered in the text (Secs.~\ref{s:V_R01_res} and
\ref{s:V_LP06_res}), and peak luminosity for a homogeneous
sample of 41 {\it Swift}/BAT GRBs.}
 \label{table:results}
 \begin{tabular}{llrrrrr}
 \hline
 \hline
 \noalign{\smallskip}
 GRB    &  $z$          &  $T_{f=0.45}$ &   $V_{\rm R}$                         &  Peak Lum. $L^{\rm (a)}$  & $V_{\rm LP}$    & References for $z$  \\
        &               &  (s)          &                                       &  $10^{50}$ erg s$^{-1}$   &  &        \\
 \noalign{\smallskip}
 \hline
 \noalign{\smallskip}
050126&	 1.29	&	$12.29$	&	$-0.005_{-0.040}^{+0.041}$ &	$14.73\pm8.53$ 	&	$-0.0506\pm0.0893$ &	\cite{Berger2005:050126redshift}	\\
050223&	$0.5915$ &	$9.73$	&	$0.084_{-0.053}^{+0.053}$ &	$1.47\pm0.65$ 	&	$-0.0986\pm0.0805$ &	\cite{Berger2005:050223redshift}	\\
050315&	$1.949$ &	$24.96$	&	$0.081_{-0.012}^{+0.012}$ &	$29.44\pm4.97$ 	&	$-0.0026\pm0.0063$ &	\cite{Kelson2005:050315redshift}	\\
050319&	$3.24$  &	$12.54$	&	$0.285_{-0.044}^{+0.044}$ &	$90.91\pm14.00$ &	$0.0046\pm0.0034$ &	\cite{Fynbo2005:050319redshift}	\\
050401&	$2.9$   &	$4.80$	&	$0.175_{-0.021}^{+0.020}$ &	$1405.1\pm165.3$&	$0.0176\pm0.0035$ &	\cite{Fynbo2005:050401redshift}	\\
050416A&$0.6535$ &	$1.47$	&	$0.185_{-0.092}^{+0.092}$ &	$0.85\pm0.25$ 	&	$-0.0083\pm0.0064$ &	\cite{Cenko2005:050416Aredshift}	\\
050505&	$4.27$  &	$10.50$	&	$0.175_{-0.036}^{+0.036}$ &	$369.00\pm42.00$&	$-0.0060\pm0.0163$ &	\cite{Berger2005:050505redshift}	\\
050525A&$0.606$ &	$2.62$	&	$0.096_{-0.004}^{+0.005}$ &	$57.11\pm15.30$ &	$0.0022\pm0.0002$ &	\cite{Foley2005:050525Aredshift}	\\
050603&	$2.821$ &	$2.43$	&	$0.286_{-0.030}^{+0.031}$ &	$2706.5\pm1470.0$ &	$0.0090\pm0.0014$ &	\cite{Berger2005:050603redshift}	\\
050730&	$3.967$ &	$54.72$	&	$0.063_{-0.024}^{+0.024}$ &	$87.14\pm19.24$ &	$-0.0404\pm0.0284$ &	\cite{Chen2005:050730redshift}	\\
050803&	$0.422$ &	$20.48$	&	$0.094_{-0.029}^{+0.029}$ &	$1.91\pm0.56$ 	&	$-0.0007\pm0.0072$ &	\cite{Bloom2005:050803redshift}	\\
050814&	$5.3$   &	$54$	&	-- &	$196.78\pm64.28$&	$-0.0118\pm0.0083$ &	\cite{Jakobsson2006:050814redshift}	\\
050824&	 $0.83$ &	$12$	&	-- &	$0.202\pm0.0145$&	$-0.3938\pm0.2506$ &	\cite{Fynbo2005:050824redshift}	\\
050908&	$3.35$ &	$6.40$	&	$-0.012_{-0.032}^{+0.032}$ &	$73.00\pm15.00$ &	$-0.0373\pm0.0324$ &	\cite{Fugazza2005:050908redshift}	\\
050922C&$2.198$ &	$1.34$	&	$0.026_{-0.005}^{+0.005}$ &	$443.05\pm21.10$&	$0.0055\pm0.0018$ &	\cite{Jakobsson2005:050922Credshift}	\\
051016B&$0.9364$ &	$3.26$	&	$0.272_{-0.086}^{+0.094}$ &	$4.85\pm1.19$ 	&	$-0.0092\pm0.0055$ &	\cite{Soderberg2005:051016Bredshift}	\\
051109A&$2.346$ &	$9.79$	&	$0.154_{-0.069}^{+0.076}$ &	$274.18\pm44.50$&	$-0.0167\pm0.0123$ &	\cite{Quimby2005:051109Aredshift}	\\
051111&	$1.55$  &	$11.20$	&	$0.026_{-0.006}^{+0.005}$ &	$103.88\pm12.18$&	$-0.0009\pm0.0022$ &	\cite{Hill2005:051111redshift}	\\
060115&	$3.53$  &	$27.65$	&	$0.120_{-0.024}^{+0.031}$ &	$115.56\pm17.22$&	$-0.0140\pm0.0089$ &	\cite{Piranomonte2006:060115redshift}	\\
060206&	$4.048$ &	$3.84$	&	$0.054_{-0.022}^{+0.022}$ &	$444.52\pm20.18$&	$-0.0038\pm0.0022$ &	\cite{Fynbo2006:060206redshift}	\\
060210&	$3.91$  &	$40.77$	&	$0.203_{-0.022}^{+0.021}$ &	$542.42\pm40.56$&	$0.0038\pm0.0025$ &	\cite{Cucchiara2006:060210redshift}	\\
060223A&$4.41$  &	$6.72$	&	$0.106_{-0.036}^{+0.037}$ &	$244.49\pm24.72$&	$-0.0174\pm0.0148$ &	\cite{Berger2006:060223redshift}	\\
060418&	$1.49$ &	$16.70$	&	$0.184_{-0.009}^{+0.009}$ &	$131.65\pm9.89$ &	$0.0053\pm0.0006$ &	\cite{Dupree2006:060418redshift_initial}\\
060502A&$1.51$  &	$9.22$	&	$0.006_{-0.005}^{+0.006}$ &	$87.44\pm15.11$ &	$-0.0130\pm0.0075$ &	\cite{Cucchiara2006:060502redshift}	\\
060510B&$4.9$   &	$92.16$	&	$0.105_{-0.015}^{+0.014}$ &	$143.84\pm22.46$&	$0.0013\pm0.0220$ &	\cite{Price2006:060510Bredshift}	\\
060512&	$0.4428$ &	$3.46$	&	$0.058_{-0.080}^{+0.077}$ &	$0.15\pm0.10$ 	&	$-0.2220\pm0.0842$ &	\cite{Bloom2006:060512redshift}	\\
060522&	$5.11$  &	$22.08$	&	$0.083_{-0.051}^{+0.049}$ &	$90.26\pm25.11$ &	$-0.0197\pm0.0166$ &	\cite{Cenko2006:060522redshift}	\\
060526&	$3.21$  &	$17.02$	&	$0.298_{-0.044}^{+0.047}$ &	$189.93\pm20.05$&	$0.0003\pm0.0011$ &	\cite{Berger2006:060526redshift}	\\
060604&	$2.68$  &	$8.96$	&	$0.189_{-0.130}^{+0.131}$ &	$17.42\pm5.46$ 	&	$-0.9493\pm0.5234$ &	\cite{Castro-Tirado2006:060604redshift}	\\
060605&	$3.7$   &	$19.01$	&	$0.097_{-0.062}^{+0.061}$ &	$99.03\pm20.89$ &	$-0.0657\pm0.0259$ &	\cite{Still2006:060605redshift}	\\
060607&$3.082$ &	$22.08$	&	$0.171_{-0.022}^{+0.018}$ &	$164.79\pm16.27$&	$-0.0010\pm0.0016$ &	\cite{Ledoux2006:060607redshift}	\\
060614&	$0.125$ &	$24.90$	&	$0.274_{-0.010}^{+0.010}$ &	$0.80\pm0.11$ 	&	$0.0049\pm0.0006$ &	\cite{Fugazza2006:060614redshift}	\\
060707&	$3.43$  &	$20.35$	&	$0.096_{-0.046}^{+0.044}$ &	$98.96\pm21.02$ &	$-0.0029\pm0.0297$ &	\cite{Jakobsson2006:060707redshift}	\\
060714&	$2.71$  &	$22.40$	&	$0.180_{-0.021}^{+0.021}$ &	$88.78\pm10.53$ &	$-0.0021\pm0.0079$ &	\cite{Jakobsson2006:060714redshift}	\\
060729&	$0.54$  &	$26.62$	&	$0.165_{-0.064}^{+0.064}$ &	$0.49\pm0.35$ 	&	$-0.0036\pm0.0309$ &	\cite{Thoene2006:060729redshift}	\\
060904B&$0.703$ &	$6.91$	&	$0.109_{-0.035}^{+0.027}$ &	$17.16\pm3.05$ 	&	$0.0003\pm0.0008$ &	\cite{Fugazza2006:060904Bredshift}	\\
060908&	$2.43$  &	$5.76$	&	$0.106_{-0.014}^{+0.011}$ &	$280.00\pm24.00$&	$0.0021\pm0.0036$ &	\cite{Rol2006:060908redshift}	\\
060912A&$0.937$ &	$1.28$	&	$0.025_{-0.009}^{+0.012}$ &	$46.20\pm4.00$ 	&	$-0.0011\pm0.0015$ &	\cite{Jakobsson2006:060912Aredshift}	\\
060926&	$3.208$ &	$3.07$	&	$0.059_{-0.033}^{+0.034}$ &	$55.00\pm9.00$ 	&	$0.0122\pm0.0182$ &	\cite{D'Elia2006:060926redshift}	\\
060927&	$5.6$   &	$3.84$	&	$0.155_{-0.021}^{+0.022}$ &	$984.00\pm590.00$&	$0.0125\pm0.0023$ &	\cite{Fynbo2006:060927redshift}	\\
061007&	$1.262$ &	$17.54$	&	$0.123_{-0.002}^{+0.002}$ &	$675.16\pm28.51$s&	$0.0117\pm0.0005$ &	\cite{Jakobsson2006:061007redshift}	\\

  \noalign{\smallskip}
  \hline
  \end{tabular}
  \end{center}
  \begin{list}{}{}
  \item[$^{\mathrm{a}}$]{Isotropic-equivalent peak luminosity in $10^{50}$~erg s$^{-1}$ 
  in the rest-frame 100--1000~keV band, for peak fluxes measured on a 1-s time-scale, 
  $H_0 = 65$ km s$^{-1}$ Mpc$^{-1}$, $\Omega_m = 0.3$, and $\Omega_{\Lambda} = 0.7$.}
  \end{list}
  \end{table*}

\subsection[]{R01 definition}
\label{s:V_R01_res}
\begin{figure*}
\begin{center}
\centerline{\includegraphics[width=15cm]{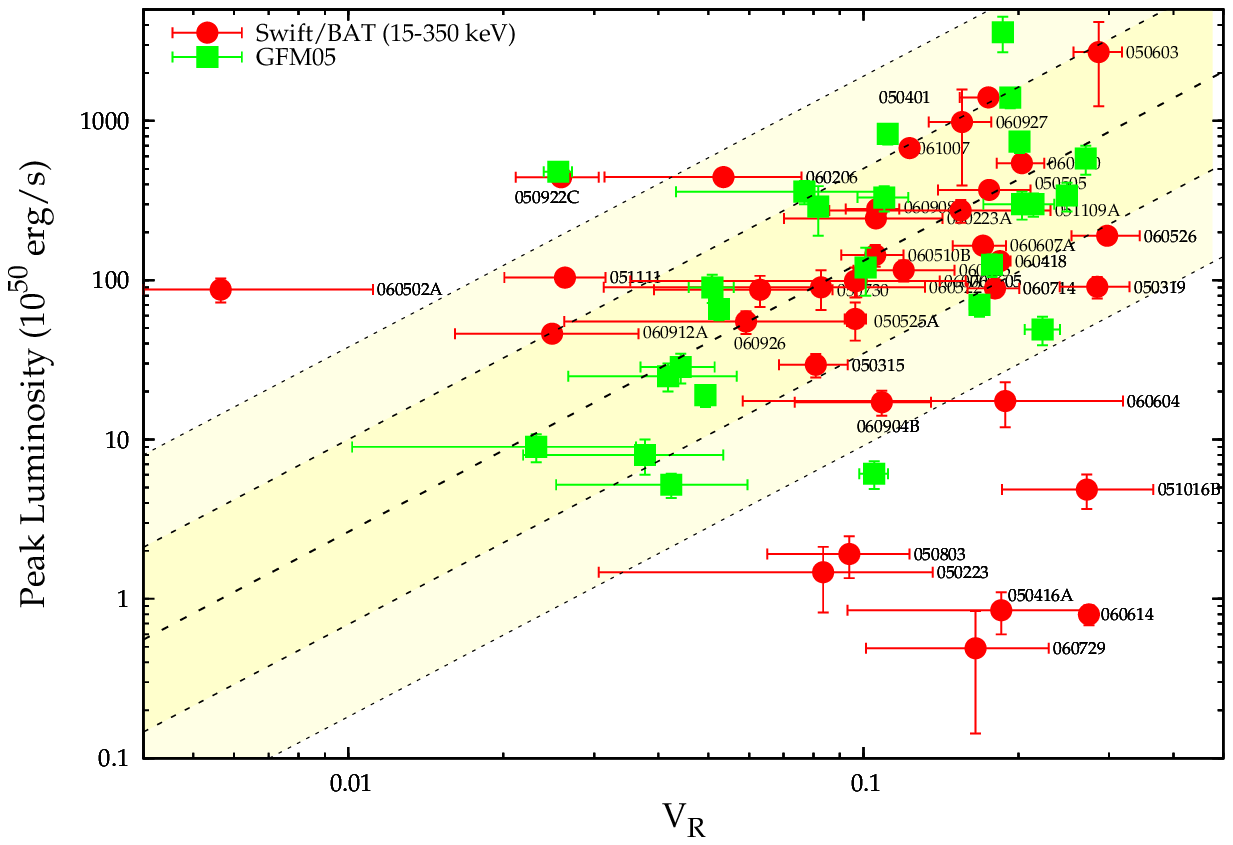}}
\caption{Variability $V_{\rm R}$ vs. peak luminosity $L$ for a sample of 36 long bursts detected
by {\it Swift}/BAT (circles) according to the definition of variability by \citet{Reichart2001:luminosity}.
For comparison we show 25 GRBs (squares) from \citet{Guidorzi2005:var_peaklum}.
The shaded areas show the 1- and 2-$\sigma$ regions around the best-fit power law obtained by \citet{Guidorzi2006a}
with the D'Agostini method, with a slope of 1.7.}
\label{f:reichart_varlum}
\end{center}
\end{figure*}

Significant values of $V_{\rm R}$ have been obtained for 36 GRBs shown in Fig.~\ref{f:reichart_varlum}
(circles).
In the remaining 5 cases this was not possible for different reasons. For GRB~050814 and GRB~050824 we
could not find any binning matching the requirements mentioned above.
While for GRB~050126, GRB~050908 and GRB~060512 $V_{\rm R}$ turned out to be consistent with zero
within uncertainties.
%
%
Figure~\ref{f:reichart_varlum} also shows the sample of 26~GRBs of GFM05 (squares): the underluminous
GRB~980425, which belongs to the GFM05 sample, is not shown because of scale compression reasons;
moreover, its uncertainty on $V_{\rm R}$ is relatively large.

We do not show the values GFM05 estimated
for six {\it Swift}/BAT bursts in common with our sample.
Except for the case of GRB~050319, our values of $V_{\rm R}$
for the other 5 GRBs are broadly consistent with those of GFM05, some differences being due to a different
energy band choice (see above).
In general, we note that our $T_f$ are systematically somewhat higher
than those of GFM05: this is so because we included low-energy bands, in which GRBs
are known to last longer. In addition, we know that in some cases $V_{\rm R}$ has a strong
dependence on the energy band (GFM05), although the definition of $V_{\rm R}$ by R01 was originally
thought to account for the narrowing of pulses at higher energies \citep{Fenimore95,Norris96}.
In the case of GRB~050319 we measured $V_{\rm R}=0.285\pm0.044$, while GFM05 obtained
$V_{\rm R}=0.06\pm0.03$. The inconsistency is due to the fact that the original
event file, available at the time and used by GFM05 to extract the light curve, was missing
the first sequence of impulses well before the trigger time. Therefore, we consider the value
reported in this paper as the correct one.

We tested the existence of the $V/L$ correlation over a number of different GRB data sets.
Our sample of 36 BAT GRBs shows no significant correlation according to Pearson's, Spearman's
and Kendall's coefficients, whose corresponding no-correlation probabilities are 72\%, 51\% and
37\%, respectively. However, from Fig.~\ref{f:reichart_varlum} we note that in the region of high
$V_{\rm R}$ and low $L$, rather unexplored by previous data sets (R01; GFM05), there are
six GRBs: 050223, 050416A, 050803, 051016B, 060614, 060729.
If one selects the BAT GRBs from our sample with $L_{50}>5$,
the resulting sample of 30 GRBs shows a significant
improvement of the $V/L$ correlation: the probability of no correlation becomes 16\%, 5.1\% and 3.1\%,
respectively.
Likewise, if we merge the two samples (GFM05's and ours) we obtain similar results:
when the 7 bursts with $L_{50}<5$ are
taken out from the total sample of 62 GRBs, the correlation becomes significant with a no-correlation
probability of $\sim2\times10^{-4}$ according to the non-parametric tests.

Finally, we calculated $V_{\rm R}$ in the 25--350~keV energy band, i.e. ignoring the
lowest energy channel 15--25~keV, of the six low-luminosity outliers.
The aim was to establish the importance of the low-energy channel contribution to
the resulting $V_{\rm R}$, especially when compared with the results of
GFM05, whose low-energy threshold was 40~keV. We found that in all cases $V_{\rm R}$
resulted systematically higher, although still compatible within uncertainties.
The only case in which $V_{\rm R}$ in the 25--350~keV was significantly higher than
for the whole band was 060614 due to its small statistical uncertainty.
This corroborates the nature of outliers of the six GRBs considered: we
can rule out that their high values of $V_{\rm R}$ are due to the presence
of the low-energy photons not considered by previous data sets.

\subsection[]{LP06 definition}
\label{s:V_LP06_res}
\begin{figure*}
\begin{center}
\centerline{\includegraphics[width=15cm]{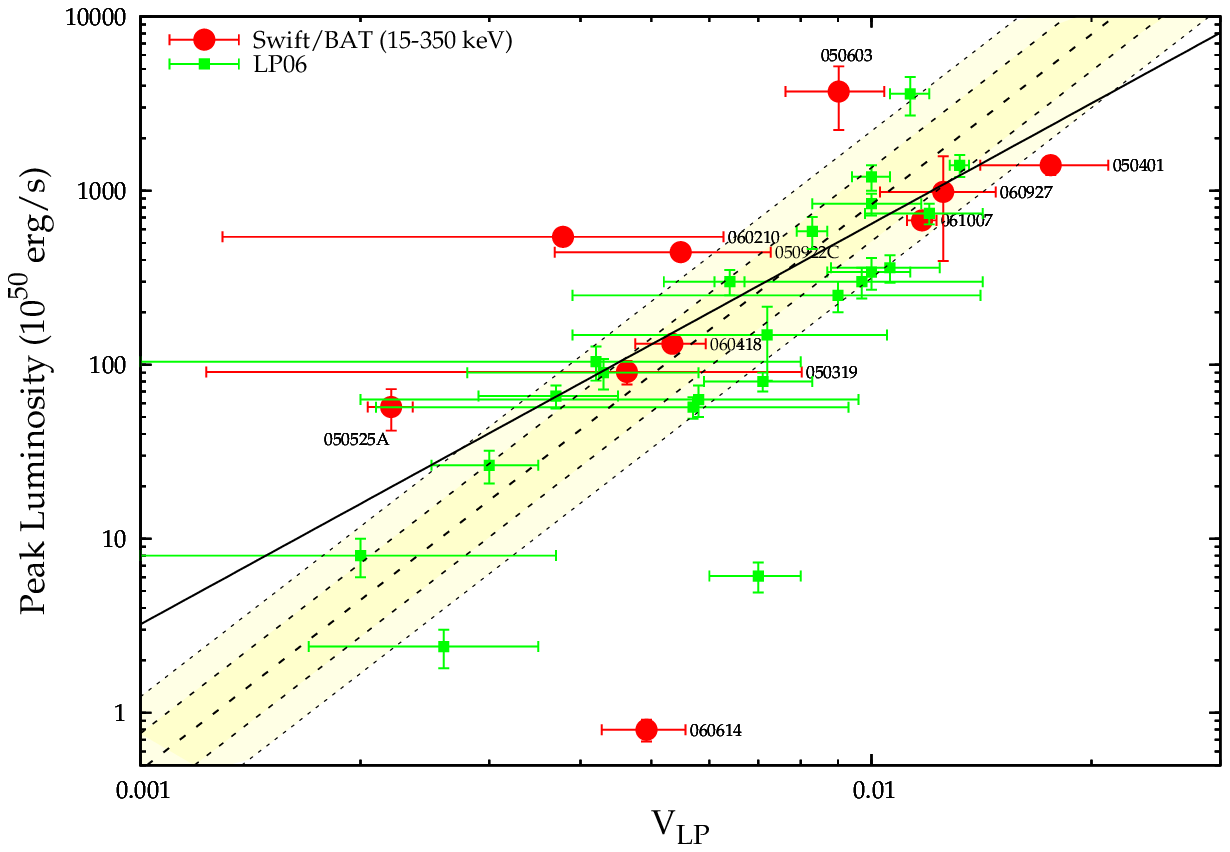}}
\caption{Variability $V_{\rm LP}$ vs. peak luminosity $L$ for a sample of 10 long bursts detected
by {\it Swift}/BAT (circles) according to the definition of variability by \citet{Li2006:var_lum}.
For comparison we show 22 GRBs with significant $V_{\rm LP}$ (squares) from \citet{Li2006:var_lum}.
The shaded areas show the 1- and 2-$\sigma$ regions around the best-fit power law obtained by
\citet{Li2006:var_lum}, with a slope of 3.25. Solid line shows the best-fitting power law obtained
with all of the 10 BAT bursts shown here, but GRB~060614; its slope is 2.3
(see Sec.~\ref{s:V_LP06_res}).}
\label{f:lp_varlum}
\end{center}
\end{figure*}

Significant values of $V_{\rm LP}$ have been obtained only for 10 GRBs shown in Fig.~\ref{f:lp_varlum}.
In the remaining 31 cases the variability resulted consistent with zero within uncertainties (see
Table~\ref{table:results}).
%
%
Despite the small number of GRBs with significant $V_{\rm LP}$, the correlation appears to
be significant within 1--2\% according to the non-parametric tests: 1.1\% (Spearman)
and 1.6\% (Kendall). See Table~\ref{tab:corr_coeff} for further details.
Figure~\ref{f:lp_varlum} shows these 10 BAT GRBs as well as the sample of 22 GRBs
of LP06. Shaded areas show the 1-$\sigma$ and 2-$\sigma$ regions around the best-fitting
power law  obtained by LP06 using the {\em fitexy} routine, with a slope of $m=3.25\pm0.26$
and a $\chi^2/{\rm dof}=1.93$ (20 dof). If we ignore GRB~060614, which clearly lies far
away from any power-law correlation between $V_{\rm LP}$ and $L$, and use the same routine as
LP06, we obtain a best-fitting value of the slope of $m=2.3\pm0.17$ and
$\chi^2/{\rm dof}=8.5$ (7 dof). The $\chi^2$ is clearly too high and therefore, although
the correlation appears to be real, the description in terms of a power-law with no
sample scatter, as the usage of the routine {\em fitexy} assumes, is not acceptable.
We note that this conclusion also holds for the very same result of LP06, whose $\chi^2$
has a null hypothesis probability of 0.75\%.
 
%
%
\begin{table*}
\centering
  \caption{Correlation Coefficients for different sets of GRBs.
    }
  \label{tab:corr_coeff}
  \begin{tabular}{lccc}
\hline
Set of GRB$^{\rm (a)}$ & \multicolumn{3}{c}{Coefficient (Probability)} \\
           & Pearson's $r$ & Spearman's $r_s$ & Kendall's $\tau$\\
\hline
36 GRBs ($V_{\rm R}$ vs. $L_{50}$)    & $-0.062$ ($0.719$) & $0.115$ ($0.506$) & $0.105$ ($0.369$)\\
30 GRBs ($V_{\rm R}$ vs. $L_{50}>5$)  &  $0.261$ ($0.163$) & $0.359$ ($0.051$) & $0.278$ ($0.031$)\\
62 GRBs$^{\rm (b)}$ ($V_{\rm R}$ vs. $L_{50}$)    &  $0.190$ ($0.139$) & $0.315$ ($0.013$) & $0.231$ ($0.008$)\\
55 GRBs$^{\rm (b)}$ ($V_{\rm R}$ vs. $L_{50}>5$)  &  $0.418$ ($1.5\times10^{-3}$) & $0.476$ ($2.4\times10^{-4}$) &  $0.342$ ($2.3\times10^{-4}$)\\ 
10 GRBs ($V_{\rm LP}$ vs. $L_{50}$)   &  $0.536$ ($0.111$) & $0.758$ ($0.011$) & $0.600$ ($0.016$)\\
\hline
\end{tabular}
 \begin{list}{}{}
  \item[$^{\mathrm{a}}$ $L_{50} = L/(10^{50}~{\rm erg}~{\rm s}^{-1})$.]
  \item[$^{\mathrm{b}}$ This sample resulted from the merging of our sample with that of \citet{Guidorzi2005:var_peaklum}.]
  \end{list}
\end{table*}

\section[]{Discussion}
\label{s:disc}
Interestingly, if one ignores the 6 GRBs from our sample of {\it Swift}/BAT with low
$L$, specifically $L_{50}<5$, the remaining homogeneous sample of 30 BAT GRBs, for which
we could derive a reliable estimate of $V_{\rm R}$ in the 15--350~keV energy band,
is fully consistent in the $V_{\rm R}$-$L$ plot with those from previous detectors, 
thus confirming the existence of the $V_{\rm R}/L$ correlation.
This is remarkable, given that BAT is a different kind of $\gamma$-ray detector and
has a different energy band from that of the
{\it BeppoSAX}/GRBM, 40--700~keV, whose data mainly comprise the sample of
32 GRBs of GFM05.
Another important confirmation provided by this BAT sample is that the scatter
of the correlation originally found by R01 and GFM05, despite their alternative
descriptions of it, is not due to the combination of data from different instruments
with different effective areas, response functions, statistical noises, and energy bands,
but it is intrinsic to the correlation. In fact, for the first time
our data set represents a homogeneous sample of 36 GRBs with measured redshift
acquired with the very same detector and with the very same kind of data for each GRB.

What is new with this BAT sample is the presence of 6 (out of 36) low-luminosity
GRBs ($L_{50}<5$). If one ignores GRB~980425,
a peculiar underluminous and very nearby burst, 
from the sample of GFM05 and R01 it turns out that none of the previous GRBs
has $L_{50}<5$. This is not surprising, given the unprecedented sensitivity of BAT.
Therefore these 6 BAT GRBs allow us to test, for the first
time, whether the correlation holds for low-luminosity GRBs. Figure~\ref{f:reichart_varlum}
clearly shows that none of them lies where one might have expected from the
correlation. Instead, they exhibit relatively high values
of $V_{\rm R}$. This is proven by the correlation coefficients, in particular
the non-parametric Spearman's $r_s$ and Kendall's $\tau$, according
to which the correlation is significant (5.1\% and 3.1\% respectively)
or not, depending whether these
6 low-luminosity GRBs are excluded or not.
This is confirmed by merging our sample of BAT with that of GFM05: the correlation
is significant, provided that low-luminosity bursts are excluded (see
Table~\ref{tab:corr_coeff}).

\citet{Guidorzi2006b} have investigated the nature of the 6 BAT GRBs outliers
of the $V/L$ correlation and found strong evidence that they are also outliers
of the anti-correlation, discovered by \citet{Norris00}, between the rest-frame
temporal lag and the peak luminosity. In particular, they found that these
GRBs are characterised by a small or negligible time lags and a relatively low
luminosity. We refer the reader to the paper by \citet{Guidorzi2006b} for more details.

Concerning the definition of variability, $V_{\rm LP}$, given by LP06, we
found that this still correlates with $L$, although our results differ from
those by LP06 (see Fig.~\ref{f:lp_varlum} and Table~\ref{tab:corr_coeff}).
In particular, we find the description of the correlation 
in terms of a power law with no extrinsic scatter inadequate, given the high
values of $\chi^2/{\rm dof}$ yielded by both samples, ours and LP06's.
Regarding our sample of 41 BAT GRBs, we find that, unlike the definition of
$V_{\rm R}$ by R01, the smoothing filter adopted
by LP06 in their definition of $V_{\rm LP}$ cuts off the low-frequency
variability of GRBs. This results in a selection of a smaller sample of 
GRBs with significant (high-frequency) variability: 10 vs. the 36
obtained for the R01 definition.
We note that GRB~060614 confirms its nature of outlier of the correlation, 
no matter which choice of the definition of variability we adopt 
(Fig.~\ref{f:lp_varlum}).

In general, from Table~\ref{tab:corr_coeff} we note that the Pearson linear
correlation coefficient $r$ is systematically less significant than the other
two. This supports the finding that the correlation
shows a clear scatter around the best-fitting power law.
Therefore this scatter must be taken into account properly (e.g. with the
D'Agostini method), when fitting the data (see D'Agostini 2005 and Guidorzi
et~al. 2006).\nocite{DAgostini05,Guidorzi2006a}

\subsection[]{Low-luminosity GRBs and the Amati correlation}
\label{s:epi}
We tested if the 6 low-luminosity GRBs are also outliers of the $E_{\rm p,i}$-$E_{\rm iso}$
\citep{Amati02} ($E_{\rm iso}$ is the isotropic energy released in the $1-10^4$~keV rest-frame
band) as well as of the $E_{\rm p,i}$-$L$ \citep{Yonetoku04,Ghirlanda05} correlations.
$E_{\rm p,i}=E_{\rm p} (1+z)$ is the intrinsic
peak energy of the total spectrum of a burst, where $E_{p}$ is the peak of the $\nu F(\nu)$
spectrum in the observer frame.
A correlation between temporal variability and $E_{\rm p,i}$ was originally found
by \citet{Lloyd02} for a number of bursts with pseudo-redshift derived assuming the
variability/peak luminosity correlation. 
Taking into account that $E_{\rm p,i}$ also correlates with $E_{\rm iso}$
and with $L$ (isotropic peak luminosity), we test whether the breaking of the
$V/L$ correlation in the case of these 6 bursts is explained by anomalous values of $E_{\rm p,i}$.

For two bursts, XRF~050416A \citep{Sakamoto06} and GRB~060614 \citep{Amati07} $E_{\rm p,i}$ has already
been reported elsewhere. Both GRBs are consistent with the Amati relation. In particular, XRF~050416A
remarkably confirms it down to the XRFs region \citep{Sakamoto06}.
For the remaining four GRBs, the BAT photon spectrum can be fit with
a single power law $N(E)\propto E^{-\Gamma_{\rm BAT}}$, where $\Gamma_{\rm BAT}$ is the photon index.
In order to constrain $E_{\rm p}$, we fitted the total spectrum of
each burst with a cutoff power law by fixing the power law index $\alpha$ to the typical value of
$1.0$ and letting the break energy $E_0=E_{\rm p}/(2-\alpha)$ free to vary.
We took the lower/upper limit for $E_0$ from the 90\% confidence level interval on
one parameter: if the interval included or lay close to the lower (higher) boundary of the BAT passband,
we assumed the upper (lower) limit on $E_0$. Our results are broadly in agreement with
the empirical correlation found by \citet{Zhang07} between $E_{\rm p}$ and $\Gamma_{\rm BAT}$.

%
%
\begin{figure}
\begin{center}
\centerline{\includegraphics[width=9cm]{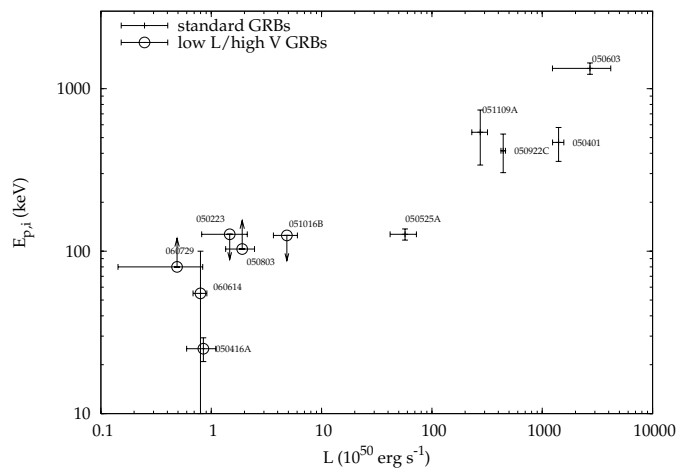}}
\caption{Peak luminosity $L$ vs. rest-frame peak energy $E_{{\rm p,i}}$ of the total energy spectrum
for 5 bursts with firm $E_{{\rm p,i}}$ measurements \citep{Amati06} and the 6 low-luminosity
($L_{50}<5$) GRBs (empty circles) of our {\it Swift}/BAT sample.}
\label{f:epi}
\end{center}
\end{figure}
%
%
%
\begin{table}
\centering
  \caption{Intrinsic peak energy $E_{{\rm p,i}}$ of the total spectrum for
the subset of 6 low-luminosity GRBs of our sample. $\Gamma_{\rm BAT}$ is the photon index of the
total photon spectrum ($N(E)\propto E^{-\Gamma_{\rm BAT}}$) when this is fit with a single
power law in the BAT energy band. Limits are given at 90\% confidence level.}
  \label{tab:epi}
  \begin{tabular}{lcrr}
\hline
GRB & $\Gamma_{\rm BAT}$ & $E_{{\rm p,i}}$ (keV) & $E_{\rm iso}$ ($10^{52}$ erg)\\\hline
050223    & $1.90\pm0.16^{\rm (a)}$ & $<127$ & $0.12\pm0.02$\\
050416A$^{\rm (b,c)}$   & --          & $25.1\pm4.2$ & $0.12\pm0.02$\\
050803    & $1.58\pm0.09$ & $>103$ & $0.20\pm0.03$ \\
051016B   & $2.13\pm0.27$ & $<125$ & $0.14\pm0.04$\\
060614$^{\rm (d)}$    & --          & $55\pm45$ & $0.25\pm0.10$\\
060729    & $1.62\pm0.18$ & $>80$ & $0.27\pm0.05$ \\
\hline
\hline
\end{tabular}
 \begin{list}{}{}
  \item[$^{\mathrm{a}}$ In agreement with \citet{Page05}.]
  \item[$^{\mathrm{b}}$ from \citet{Amati06}.]
  \item[$^{\mathrm{c}}$ from \citet{Sakamoto06}.]
  \item[$^{\mathrm{d}}$ from \citet{Amati07}.]
  \end{list}
\end{table}

Results are reported in Table~\ref{tab:epi}. All of the 6 bursts (or their limits) turned out
to lie in the 2 sigma region of the Amati relation (see Amati 2006).

%
%
\begin{figure}
\begin{center}
\centerline{\includegraphics[width=9cm]{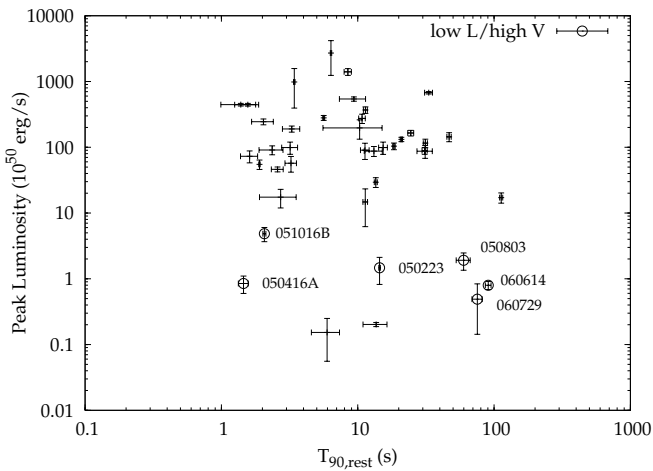}}
\caption{Rest-frame duration $T_{90,{\rm rest}}$ vs. peak luminosity for all the 41 {\it Swift}/BAT
bursts reported in Table~\ref{table:results}.
Empty circles show the 6 low-luminosity ($L_{50}<5$) high-variability GRBs.}
\label{f:t90_vs_L}
\end{center}
\end{figure}

We also found that the two bursts with firm $E_{\rm p,i}$ as well
as two with upper limits are consistent with the $E_{\rm p,i}$-$L$ correlation, while the remaining
two lower limits on $E_{\rm p,i}$ for GRB~050803 and GRB~060729 are not, as shown in Fig.~\ref{f:epi}.
The better consistency with the $E_{\rm p,i}$-$E_{\rm iso}$ than with the $E_{\rm p,i}$-$L$
correlation can be explained with the fact that $E_{\rm p,i}$ better correlates with the time
integrated released energy, as proven also by the scatter of the correlation between $L$
and $E_{\rm iso}$ \citep{Ghirlanda05}.

We also tested whether the duration of these events correlates with their peak luminosity.
To this aim, in Fig.~\ref{f:t90_vs_L} the rest-frame $T_{90,{\rm rest}}=T_{90}/(1+z)$ is plotted
against $L_{50}$ for the entire sample of 41 {\it Swift}/BAT ~GRBs considered.
$T_{90}$ is the time interval collecting from 5\% to 95\% of the total fluence in the
observer frame.
For each burst we used the value published by the BAT team in the refined GCN circulars.
Empty circles correspond to the 6 low-luminosity GRBs with a significant measure of variability.
Apparently there is no hint for correlation and also no evidence for a different behaviour
of the 6 low-luminosity GRBs with respect to the others.
The result does not change in essence when we replace $T_{90,{\rm rest}}$ with $T_{90}$.

We conclude that the fact that the variability of these 6 low-luminosity high-variability GRBs
does not correlate with the peak luminosity is not connected with their $E_{\rm p,i}$, which
correlates with $E_{\rm iso}$ as almost all of the long GRBs with known redshift \citep{Amati06}.

\section[]{Conclusions}
\label{s:conc}
We tested the variability/peak luminosity ($V/L$) correlation with a homogeneous
sample of 36 GRBs detected with {\em Swift}/BAT in the 15--350~keV
energy band with firm redshift. We adopted two different definitions
of variability: that by \nocite{Reichart2001:luminosity}
Reichart et al. (2001; $V_{\rm R}$) and
that by Li \& Paczy{\'n}ski\nocite{Li2006:var_lum} (2006; $V_{\rm LP}$),
which differs from the former for a different smoothing filter.
The most interesting results have been derived with $V_{\rm R}$.
If we consider only the
GRBs with peak luminosity $L$ comparable with those of previous
samples, i.e. $L>5\times10^{50}$~erg~s$^{-1}$ in the
rest-frame 100--1000~keV energy band, we confirm the correlation
and its intrinsic dispersion around the best-fitting power law
obtained by \citet{Guidorzi2006a}: $m=1.7\pm0.4$ ($L\propto\,V^{m}$)
and $\sigma_{\log{L}}=0.58_{-0.12}^{+0.15}$.

Interestingly, all of the 6 low-luminosity GRBs detected by
{\em Swift}/BAT turn out to be outliers to the $V/L$ correlation,
showing higher values of $V_{\rm R}$ than expected. This does not
contradict the results from previous samples of GRBs with
known redshift. Instead, we are led to conclude that the correlation
does not hold any more for low-luminosity GRBs.
We also find that these bursts are consistent with the
$E_{\rm p,i}$--$E_{\rm iso}$ correlation \citep{Amati02} and four
of them also with the $E_{\rm p,i}$--$L$ correlation \citep{Yonetoku04,
Ghirlanda05}.

Unlike the results obtained by \citet{Li2006:var_lum}, we do not
find evidence for a tighter correlation using $V_{\rm LP}$
instead of $V_{\rm R}$. Rather, fewer GRBs appear to have a significant
measure of $V_{\rm LP}$; we ascribe this to the fact that the
smoothing filter adopted by \citet{Li2006:var_lum} to
construct the reference light curve with respect to which the
variability is computed, only selects high-frequency variability.

\section*{Acknowledgments}
This work is supported by ASI grant I/R/039/04 and by the
Ministry of University and Research of Italy
(PRIN 2005025417).
We thank L. Amati for independent checks in some parts
and the anonymous referee for useful comments.
We gratefully acknowledge the contributions of dozens of
members of the BAT team who built and maintain this instrument.

\appendix
\section{Statistical noise of BAT mask-tagged light curves}
\label{s:app}
We report the analysis performed on the BAT mask-tagged light curves
of the GRBs considered in this work, aimed at studying the statistical
noise. As the GRB itself is characterised by intrinsic temporal
variance which is unknown a priori, we limited to the pre- and post-burst
regions of the light curves, where the background is supposed
to be the dominant source of statistical fluctuations.
In order to make sure that we excluded the entire light curve of the GRB,
we binned it spanning very different integration times (from 64~ms to
32~s) and checked that no trend in the residuals was visible.

Let $r_i$ and $\sigma_{r_i}$ be the count rate and its uncertainty,
respectively, of the $i$-th bin of a 64-ms mask-tagged BAT light curve.
This light curve is the result of the BAT standard pipeline already
summarised in Sec.~\ref{s:sample} (see also Barthelmy et al. 2005).
Uncertainties $\sigma_{r_i}$ ($i=1,\ldots,N$, where $N$ is the total
number of bins of the selected portion of light curve) are calculated by
propagation of errors, starting from the raw counts assumed to
be affected by purely Poissonian noise through the ftool {\em batbinevt}.

We tested the following null hypothesis: each $r_i$ is a single
realisation of a normal random variable with null expected value and
sigma equal to $\sigma_{r_i}$: $N(0,\sigma_{r_i})$.
Little can be inferred on a random variable from a single realisation.
However, as long as this hypothesis is true, the various $r_i/\sigma_{r_i}$
($i=1,\ldots,N$) can be seen as different realisations of the same random variable,
$\mathbf{r_{\rm n}}$, hereafter called ``normalized rate'', which has
a standard normal density: $N(0,1)$.

So we studied the observed distribution of $\mathbf{r_{\rm n}}$ for each single
light curve removed of the GRB profile. We fitted the observed
distribution with a Gaussian $N(\mu,\sigma)$.

In particular, we are interested in constraining the possible presence
of any additional source of statistical noise (e.g. instrumental) to
the Poissonian one.

More generally, should the various $r_i$ fluctuate more than $\sigma_{r_i}$,
so that the true variance is $(1+f_{\rm np})\,\sigma_{r_i}^2$, where
$f_{\rm np}$ is the fraction of additional non-Poissonian variance,
the resulting $\sigma$ should be greater than unity. More precisely,
we should find $\sigma^2=(1+f_{\rm np})$.

Therefore we fitted the observed distribution of $\mathbf{r_{\rm n}}$ with
$N(\mu,\sigma)$, first by imposing $\sigma=1$. In every case we found
acceptable $\chi^2$ values, confirming that no evidence for additional
noise has been found.

In particular, we were interested in setting a limit to $f_{\rm np}$
with a given confidence level. Following Papoulis \& Pillai (2002; p. 313--314),
\nocite{Papoulis02} in the case of unknown $\mu$ we used the sample variance $s^2$
defined as:
\begin{equation}
s^2 = \frac{1}{N-1}\quad \sum_{i=1}^{N}\ (r_{{\rm n},i}-\overline{r_{\rm n}})^2
\end{equation}
where $r_{{\rm n},i}=r_i/\sigma_{r_i}$ is the single realisation of
$\mathbf{r_{\rm n}}$ and $\overline{r_{\rm n}}$ is the mean value.
The random variable $(N-1)\mathbf{s}^2/\sigma^2$ follows a $\chi^2(N-1)$
distribution, so that we can constrain $\sigma^2$, i.e. $(1+f_{\rm np})$,
through the following:
\begin{equation}
1 + f_{\rm np} = \sigma^2 < \frac{(N-1)\ s^2}{\chi^2_{\delta/2}(N-1)}
\end{equation}
at $(1-\delta)$ confidence level; $\chi^2_u(n)$ is the $u$ percentile of the
$\chi^2(n)$ distribution.
In most cases $N$ was big enough ($>10^3$) to ensure the following approximation:
\begin{equation}
f_{\rm np} < s^2\ \Big[\ 1 + z_{1-\delta/2}\,\sqrt{\frac{2}{(N-1)}}\ \Big] \ - \ 1
\end{equation}
where $z_u$ is the $u$ percentile of the standard normal density.

\begin{figure}
\begin{center}
\centerline{\includegraphics[width=8cm]{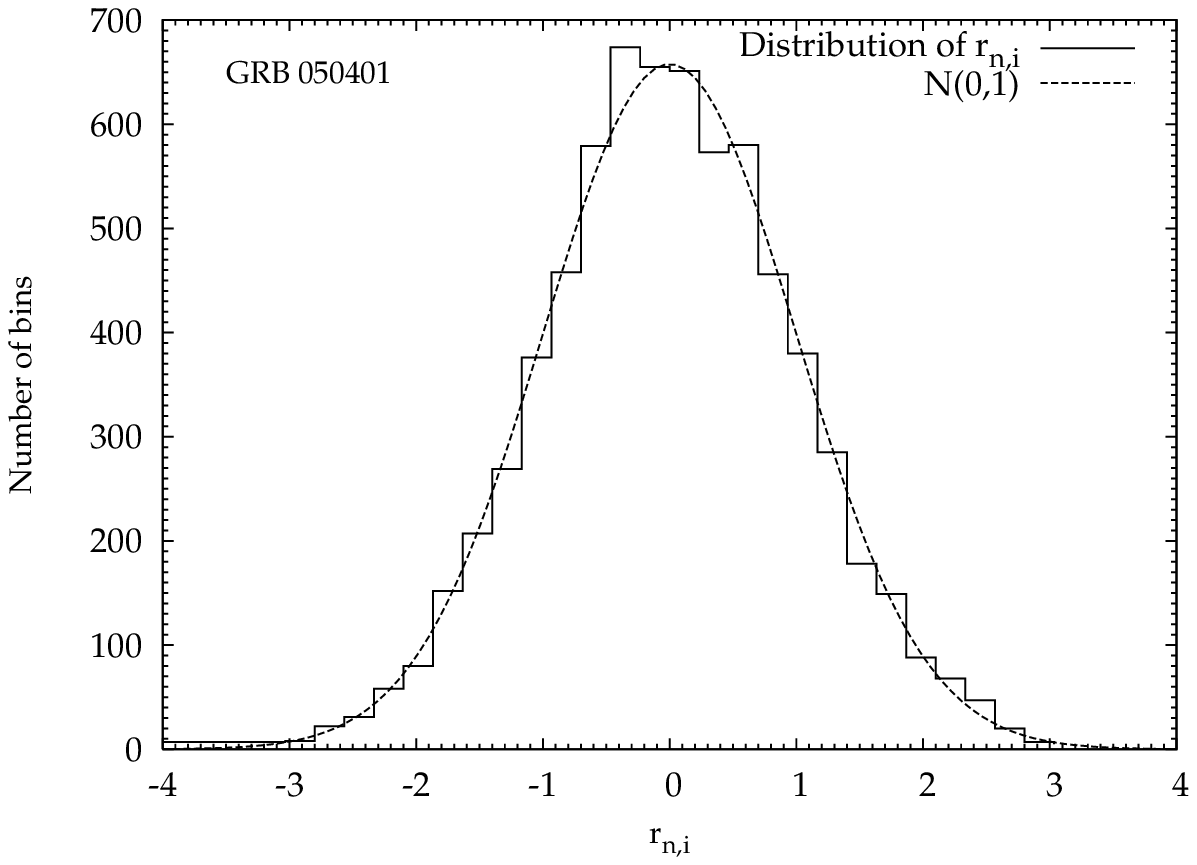}}
\caption{Distribution of the normalized rates $r_{{\rm n},i}$ in the
case of GRB~050401. The total number of bins is $N=7065$. The distribution
is well fit with $N(0,1)$.}
\label{f:distr_resid}
\end{center}
\end{figure}

We show the example of GRB~050401. The distribution of $r_{{\rm n},i}$,
$N=7065$, can be fit with a Gaussian $N(0,1)$, 
$\chi^2/{\rm dof}=32.2/28$, as shown in Fig.~\ref{f:distr_resid}.
%
%

The sample variance resulted $s^2=0.999$. The consequent upper limit
on $f_{\rm np}$ turns out to be 2.7\% (4.3\%) at 90\% (99\%) confidence level.


\label{lastpage}

\end{document}